%% file: note.tex
\documentclass[11pt]{article}
\usepackage{graphicx}
\usepackage{amsmath}

\newcommand{\BABARPubYear}    {06}

\newcommand{\BABARConfNumber} {017}
\newcommand{\SLACPubNumber} {11991}

\input pubboard/babarsym

\newcommand{\DE}{\ensuremath{\Delta E}\xspace}
\newcommand{\mES}{\ensuremath{m_\mathrm{ES}}\xspace}
\newcommand{\mDz}{\ensuremath{m_{\Dz}}\xspace}
\newcommand{\hz}{\ensuremath{h^0}\xspace}

\newcommand{\dm}{\ensuremath{\Delta m}\xspace}
\newcommand{\dt}{\ensuremath{\Delta t}\xspace}
\newcommand{\ttag}{\ensuremath{t_\mathrm{tag}}\xspace}
\newcommand{\trec}{\ensuremath{t_\mathrm{rec}}\xspace}
\newcommand{\sigmadt}{\ensuremath{\sigma_{\Delta t}}\xspace}
\newcommand{\qoverp}{\ensuremath{\frac{q}{p}}}

\newcommand{\Abar}{\ensuremath{\overline{A}}\xspace}
\newcommand{\fbar}{\ensuremath{\overline{f}}\xspace}

\newcommand{\Abarfbar}{\ensuremath{\Abar_{\fbar}}\xspace}
\newcommand{\D}{\ensuremath{D}\xspace}

\newcommand{\etal}{{\it et al.}}

\newcommand{\ket}[1]{\ensuremath{\vert #1 \rangle}}

\newcommand{\Btag}{\ensuremath{B_{\rm{tag}}}\xspace}

\newcommand{\msqKsp}{\ensuremath{m^{2}_{\KS\pip}}\xspace}
\newcommand{\msqpm}{\ensuremath{m^{2}_{\pip\pim}}\xspace}
\newcommand{\msqKsm}{\ensuremath{m^{2}_{\KS\pim}}\xspace}
\newcommand{\sinbb}{\ensuremath{\sin2\beta}\xspace}
\newcommand{\cosbb}{\ensuremath{\cos2\beta}\xspace}
\newcommand{\abslambda}{\ensuremath{|\lambda|}\xspace}

\long\def\inst#1{\par\nobreak\kern 4pt\nobreak
    {\it #1}\par\vskip 10pt plus 3pt minus 3pt}

\begin{document}

\global\emergencystretch = .9\hsize

{\pagestyle{empty}

\begin{flushright}
\babar-CONF-\BABARPubYear/\BABARConfNumber \\
SLAC-PUB-\SLACPubNumber \\
July 2006 \\
\end{flushright}

\par\vskip 5cm

\begin{center}
\Large \boldmath Measurement of \cosbb in $\Bz\ra\D^{(*)0}\hz$ decays
with a time-dependent Dalitz plot analysis of $\Dz\ra\KS\pip\pim$
\end{center}
\bigskip

\begin{center}
\large The \babar\ Collaboration\\
\mbox{ }\\
\today
\end{center}
\bigskip \bigskip

\begin{center}
\large \bf Abstract
\end{center}
We report a preliminary measurement of \cosbb in
$\Bz\ra\D^{(*)0}\hz$ decays with a time-dependent Dalitz plot analysis of
$\Dz\ra\KS\pip\pim$, where \hz is a light neutral meson such as
$\piz$, $\eta$, $\eta'$ or $\omega$. The strong phase variation on the
Dalitz plot allows the access to the angle $\beta$ with only a
two-fold ambiguity ($\beta+\pi$). Using $311\times 10^{6}$ \BB
pairs collected at the \babar\ 
detector, we obtain
$\cosbb = 0.54 \pm 0.54 \pm 0.08 \pm 0.18$,
$\sinbb = 0.45 \pm 0.36 \pm 0.05 \pm 0.07$, and
$\abslambda = 0.975^{+0.093}_{-0.085} \pm 0.012 \pm 0.002 $,
 where the first errors are
statistical, the second are experimental systematic uncertainties, and
the third are the signal Dalitz model  uncertainties.
This measurement
favors the solution of $\beta= 22^\circ$ over $68^\circ$  at an $87\%$
confidence level.

\vfill
\begin{center}

Submitted to the 33$^{\rm rd}$ International Conference on High-Energy Physics, ICHEP 06,\\
26 July---2 August 2006, Moscow, Russia.

\end{center}

\vspace{1.0cm}
\begin{center}
{\em Stanford Linear Accelerator Center, Stanford University, 
Stanford, CA 94309} \\ \vspace{0.1cm}\hrule\vspace{0.1cm}
Work supported in part by Department of Energy contract DE-AC03-76SF00515.
\end{center}

\newpage
} 

%
%
\input pubboard/authors_ICHEP2006.tex

\section{INTRODUCTION}
\label{sec:Introduction}

The Standard Model of electroweak interactions describes charge-parity
(\CP) violation as a consequence of an irreducible phase in the
three-generation Cabibbo-Kobayashi-Maskawa (CKM) quark-mixing
matrix~\cite{ref:CKM}. In this framework, the \CP parameter $\sin 2\beta$
can be measured by examining the proper-time distribution of neutral
\B decays to \CP eigenstates. This parameter has been measured with a
high precision by the \B-factories using final states containing a
charmonium and a neutral kaon~\cite{BabarBelleS2B}. The current average
from \B-factories is $\sin2\beta= 0.685\pm 0.032$~\cite{HFAG}, which leads to a
four-fold solution 
of the angle $\beta= 22\degrees$, $68\degrees$,
($22\degrees+180\degrees$), and ($68\degrees+180\degrees$).
The ambiguity can leave possible new physics undetected even with very
high precision measurements of $\sin2\beta$. 

Analyses have been attempted to resolve the ($\beta$, $\pi/2-\beta$)
ambiguity using a time-dependent angular analysis in $\Bz\ra J/\psi
\Kstarz (\KS\piz)$ decays~\cite{ref:C2B}. 
In this analysis we use
a new method 
proposed by Bondar \etal~\cite{Bondar:2005gk}, which uses
$\Bz\ra\D^{(*)0}\hz$ decays with 
a time-dependent Dalitz plot analysis of $\Dz\ra\KS\pip\pim$, where \hz is
a light neutral meson such as \piz, $\eta^{(\prime)}$, and
$\omega$. This method takes advantage of the varying strong phase on
the $\Dz\ra\KS\pip\pim$ Dalitz plot to resolve the ambiguity of the
phase $2\beta$ from the \sinbb measurements alone.
The Belle Collaboration has recently reported a measurement using this
technique~\cite{Krokovny:2006sv}; they obtained $\cosbb=
1.87^{+0.40+0.22}_{-0.53-0.32}$ and determined the sign of \cosbb to
be positive at a 98.3\% confidence level.

\begin{figure}[htb]
\begin{center}
\includegraphics[width=0.4\textwidth]{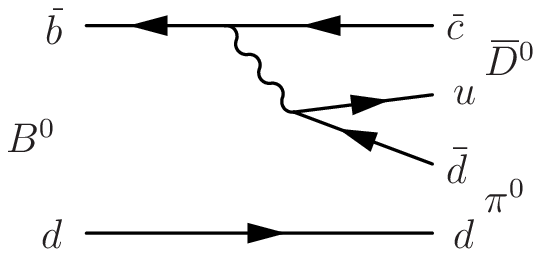}
\includegraphics[width=0.4\textwidth]{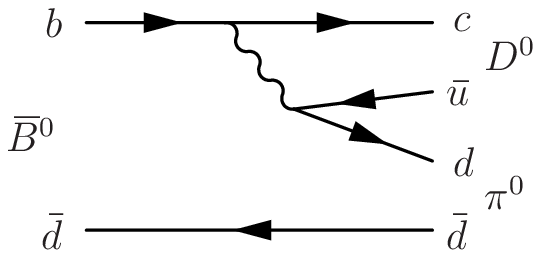}
\end{center}
\caption{Leading diagrams for $\Bz\ra\Dz\hz$ decays. }
\label{fig:diag1}
\end{figure}

The leading order diagram of $\Bz\ra\D^{(*)0}\hz$ is color-suppressed,
as shown in Figure~\ref{fig:diag1}. The next order diagram  is
suppressed by ${\cal O}(\sin^2\theta_\mathrm{Cabibbo})$.
There are no penguin diagram contributions. 
A sizable new physics effect due to supersymmetry without R-parity is
possible in $\b\ra\c\ubar\d$ decays, while the Standard Model
uncertainty is relatively small~\cite{Grossman:1996ke}.
Interference between $\Bz\ra\Dzb\hz$ and $\Bz\ra\Bzb\ra\Dz\hz$ via
mixing occurs when \Dz and \Dzb decay to a common final state, such as 
$\KS\pip\pim$.

Assuming \CPT symmetry is conserved and the decay rate difference
$\Delta \Gamma$ is negligible, the time evolution function for
a state that is known to be a \Bz at a time $t=\ttag$ can be expressed
as
\begin{equation}
\ket{\Bz_\mathrm{phys}(t)}= e^{-\Gamma \dt/2} \Big[
\ket{\Bz} \cos(\dm\dt/2) + i\qoverp \ket{\Bzb} \sin(\dm\dt/2) \Big] \;,
\end{equation}
where $\Gamma$ is the average decay rate of the two mass eigenstates of
\Bz meson, \dm is the
mixing frequency, $\dt = \trec-\ttag$ is the time difference
between \Bz decay time \trec and \ttag, and $q/p$ is the ratio of
$\ket{\Bzb}$ and $\ket{\Bz}$ coefficients in \Bz mass
eigenstates. Neglecting \CP violation in \Bz mixing, we assume
$|q/p|=1$. 
Expressing the decay amplitude of the decay chain
$\Bz\ra\Dzb\hz\ra[\KS\pip\pim]\hz$ as $A_f = AA_{\Dbar}$ and similarly
for \Bzb as $\Abarfbar = \Abar A_D$, the decay probability of a
neutral \B meson in an $\Upsilon(4S)$ system can be shown to be
\begin{eqnarray}
{\cal P}_\pm = \frac{1}{2} e^{-\Gamma \dt}|A|^2
\cdot \Big[ (|A_{\Dbar}|^2 + \abslambda^2 |A_D|^2) & \mp &
 (|A_{\Dbar}|^2 - \abslambda^2|A_D|^2)\cos(\dm \dt) \notag \\
& \pm &
2\abslambda \eta_{\hz}(-1)^L \mathrm{Im}\Big( e^{-2i\beta}
 A_D A^*_{\Dbar}\Big) \sin(\dm\dt)\Big] \;,
\label{eq:Probpm}
\end{eqnarray}
where the upper (lower) sign is for \Bz(\Bzb)-tagged events,
$\lambda= \frac{q\Abarfbar}{p A_f}$, $-2i\beta$ is the phase of $q/p$,
$\eta_{\hz}$ is the \CP eigenvalue of \hz, and $L$ is the orbital
angular momentum in the $D^{(*)0}\hz$ system. In the case of
$\Bz\ra\Dstarz\hz$ (where \hz is a pseudoscalar)
with $\Dstarz\ra\Dz\piz$, $L=1$ and two additional factors need to be
considered: the angular momentum in $\Dstarz\ra\Dz\piz$ ($L^\prime=1$)
and the \CP eigenvalue of the soft \piz from \Dstarz decay
($\eta_{\piz}=-1$)~\cite{Bondar:2004bi}.  
The decay amplitudes $A_D$ and $A_{\Dbar}$ can be expressed as a
function of two Lorentz invariant variables
$m_+^2\equiv (p_{\KS}+p_{\pip})^2$ and
$m_-^2\equiv (p_{\KS}+p_{\pi-})^2$. That is,
$A_D = A_D(m_+^2,m_-^2)$ and $A_{\Dbar} = A_D(m_-^2,m_+^2)$.
Here we have assumed that \CP is conserved in the \Dz decay and neglected
\Dz mixing.

In the last term of Equation(\ref{eq:Probpm}), one can rewrite
\begin{equation}
\mathrm{Im}\Big( e^{-2i\beta} A_D A^*_{\Dbar}\Big)
= \mathrm{Im}(A_D A^*_{\Dbar}) \cos 2\beta -
\mathrm{Re}(A_D A^*_{\Dbar}) \sin 2\beta \;,
\label{eq:sc2bsindmt}
\end{equation}
and treat \cosbb and \sinbb as independent parameters.

In this analysis, we use an unbinned maximum-likelihood method to fit
for \cosbb, \sinbb and \abslambda, and use a parameterized
Monte Carlo method based on the observed data to estimate the
confidence level of \cosbb being positive.

\input babar

\input selection

\input dalitzmodel

\input analysis

\input systematic

\input result


\section{ACKNOWLEDGMENTS}
\label{sec:Acknowledgments}

\input pubboard/acknowledgements

\end{document}

%% file: pubboard/authors_ICHEP2006.tex
\begin{center}
\small

The \babar\ Collaboration,
\bigskip

%
{B.~Aubert,}
{R.~Barate,}
{M.~Bona,}
{D.~Boutigny,}
{F.~Couderc,}
{Y.~Karyotakis,}
{J.~P.~Lees,}
{V.~Poireau,}
{V.~Tisserand,}
{A.~Zghiche}
\inst{Laboratoire de Physique des Particules, IN2P3/CNRS et Universit\'e de Savoie,
 F-74941 Annecy-Le-Vieux, France }
% ...

\end{center}\newpage

%% file: babar.tex
\section{THE \babar\ DETECTOR AND DATASET}
\label{sec:babar}
This analysis is based on $311\times 10^6$ \BB pairs collected on the
$\Upsilon(4S)$ resonance during 1999--2006 with the \babar\ detector
at the \pep2\ storage ring. A sample of $23\invfb$ collected at
40~\mev below the $\Upsilon(4S)$ resonance and a number of signal and
generic simulation samples based on Geant4~\cite{ref:Geant4} are also
analyzed to optimize the event selection and to study background properties.

The \babar\ detector is described in detail elsewhere~\cite{ref:babar}.
Charged-particle trajectories are measured by
a five-layer double-sided silicon vertex tracker and
a 40-layer drift chamber
located within a 1.5 T solenoidal magnetic field.
Charged hadrons are identified by combining energy-loss
information from the tracking system with the measurements from 
a ring-imaging Cherenkov detector.
Photons are detected by
a CsI(Tl) crystal electromagnetic calorimeter
with an energy resolution of $\sigma_E/E=0.023(E/\gev)^{-1/4}\oplus 0.014$.
The magnetic flux return is instrumented for muon and \KL\ identification.

%% file: selection.tex
\section{EVENT RECONSTRUCTION AND SELECTION}
\label{sec:selection}

In this analysis, we reconstruct \Bz decays to $\Dz\hz$, where $\hz =
\piz(\gaga)$, $\eta (\gaga, \pip\pim\piz)$,
$\eta^\prime(\pip\pim\eta)$, and $\omega(\pip\pim\piz)$, and
$\Bz\ra\Dstarz(\ra\Dz\piz)\hz$, where $\hz=\piz(\gaga)$ and $\eta(\gaga)$. 
The \Dz is reconstructed in $\KS\pip\pim$ mode.

A charged track must be reconstructed in the drift chamber, and, if it
does not result from a \KS decay, it must extrapolate back to within
1.5 cm of 
the nominal interaction point in the plane transverse to the beam and
10 cm along the beam. A cluster found in the calorimeter that is not
associated with a charged track is considered a photon candidate if
its shower shape is consistent with a photon and its energy is
greater than 30~\mev.

The \piz candidates are reconstructed by combining two photon
candidates with the \gaga invariant mass in the range 110--160~\mevcc if
used in $\Dstarz\ra\Dz\piz$ reconstruction, or 115--150~\mevcc if
used in $\Bz\ra D^{(*)0}\piz$ reconstruction; for the latter, each of
the two photons is required to have an energy greater than 50~\mev.
For $\eta\ra\gaga$, the photon candidates must both have an energy
greater than 100~\mev and the photon pair must have 
an invariant mass within 40~\mevcc of the nominal $\eta$
mass~\cite{ref:PDG} and have a momentum greater than 200\mevc in the
laboratory frame. If the $\eta\ra\gaga$ candidate is later used in a
$\Dstarz\eta$ candidate, the mass window is tightened to 33~\mevcc.
The $\eta$ candidate is removed if the invariant mass of one of the
photons and another photon in the rest of the event is within 6~\mevcc
of the nominal \piz mass.
For $\eta\to\pip\pim\piz$, the invariant mass of the
candidate is required to be within 9~\mevc of the nominal $\eta$
mass. An $\etapr$ candidate is formed by combining an $\eta$
candidate with two pions. The invariant mass must be within 8~\mevcc
of the nominal \etapr mass.
An $\omega$ candidate is formed by combining $\pip\pim\piz$.
The invariant mass of the three-pion candidate is required
to be within 18~\mevcc of the nominal $\omega$ mass. The \piz
candidate used in $\omega$ reconstruction is required to have a
momentum greater than 200~\mevc in the laboratory frame. 
Except for $\omega$, all \hz are fitted with their mass constrained at
the nominal value.

A \KS candidate consists of a vertexed pair of oppositely-charged
tracks with an invariant mass within 12~\mevcc of the nominal \KS mass
with a $\chi^2$ probability greater than 0.1\%. The \KS flight
distance must be greater than three times the estimated uncertainty,
and the angle between the flight direction and the vertex displacement
from the beam spot in the transverse plane must satisfy $\cos\theta >
0.992$. 

A \Dz candidate consists of a pair of oppositely-charged
tracks and a \KS candidate. The invariant mass, \mDz, must be within
60~\mevcc of the nominal \Dz mass. The \mDz resolution is
approximately 7~\mevcc. We retain the sideband for later use in the fit.
We then fit the \Dz kinematic parameters with
both \Dz and \KS constrained at their respective nominal mass. 
These \Dz candidates are combined with a \piz to form a
\Dstarz candidate. The invariant mass is required to be within
3.0 (2.8)~\mevcc of the \Dstarz nominal mass in $\Bz\ra\Dstarz\piz$
($\Dstarz\eta$) reconstruction.

Eventually we build a $\Bz$ candidate combining a $\piz$, $\eta$, $\omega$ or 
$\etapr$ with a $\Dz$ or a $\Dstarz$ candidate. We fit the \Bz decay
vertex requiring that the production vertex is consistent with the
beam spot in the 
transverse plane. The energy-substituted mass
$\mes\equiv\sqrt{(s/2+{\bf p}_0\cdot{\bf p}_B)^2/E_0^2- |{\bf p}_B|^2}$
is required to be greater than 5.2~\gevcc, where $s$ is the
squared center-of-mass (c.m.) energy,  
($E_0$, ${\bf p}_0$) and ($E_B$, ${\bf p}_B$) are the four-momentum of the
initial state $\epem$ and the $B$ candidate, respectively.
The energy difference
$\DeltaE \equiv E^*_B-E^*_{\rm beam}$, evaluated in the c.m. frame,
must be between $\pm 80~\mev$ ($\pm 40~\mev$) for events with
$\hz\ra\gaga$ ($\hz\ra\pip\pim\piz$ and $\etapr\ra\pip\pim\eta$).

The majority of the background is from $\qqbar$ continuum events. We
suppress them by requiring the normalized second order Fox-Wolfram moment
$R_2$~\cite{ref:R2} to be less than 0.5 and $|\cos\theta_T|$ less
than 0.9, where 
$\theta_T$ is the angle between the thrust of the $B$ candidate and the thrust 
of the rest of the event. We further suppress the continuum events by
a Fisher discriminant  formed from the following five variables: 
 $\cos\theta_T$,
 the \B flight angle in the c.m. frame,
 total event sphericity,
 total event thrust magnitude, and
the ratio of two moments $L_{2}/L_{0}$, where $L_{i}= \sum_j p^*_j
|\cos\theta^*_j|^i$, summed over the remaining particles $j$ in the
event, where $\theta^*_j$ and $p^*_j$  are the
angle with respect to the \Bz thrust and the momentum in the c.m. frame.
For $\Dz\omega$ events, two variables are added to take advantage of
the polarization of the $\omega$: the angle between the \B flight
direction and the normal to the three-pion plane in the
$\omega\ra\pip\pim\piz$ rest frame, and the angle between one pion in
the rest frame of the 
remaining pion pair with respect to the direction of the pion pair.
The Fisher coefficients are calculated using off-resonance data and
simulated signal event samples. The optimum selection value is
determined mode by mode by maximizing the signal yield significance
using simulated signal and generic background events. 

For $\Bz\ra D^{(*)0}\piz$, one major background source is the
color-allowed decay
$\Bp\ra\Dz\rho^+$, which has a branching fraction approximately 50
times larger. For events reconstructed as a $\Dz\piz$, the $\Dz\rho^+$
contribution peaks in \DE below the selection region and only
a small number of events is selected. However, for events
reconstructed as a $\Dstarz\piz$, the final state $\Dz\piz\piz$ is
very similar to $\Dz(\piz\pip)_{\rho^+}$. Therefore this background
has a mean \DE near zero. We veto $\Bp\ra\Dz\rho^+$ events by rejecting
$\Dstarz\piz$ candidates if the \piz candidate combined with any other 
charged pion in the event can form a $\rho^+$ candidate with an
invariant mass 
within 250~\mevcc of the nominal value, and subsequently form a \Bp
candidate by combining with the \Dz candidate. The requirements for
the \Bp candidate are  $\mES>5.27~\gevcc$ and
$|\DE|<100~\mev$.
Finally, we only retain events with decay time difference $|\dt|<
15$~ps and the estimated uncertainty $\sigmadt<3.6$~ps.
If there is more than one candidate in the event, the one with a more
signal-like Fisher discriminant is selected.


We use a two-dimensional $(\mES, \mDz)$ probability density function (PDF)
in an unbinned-maximum-likelihood fit to 
separate three types of events: (1) signal: 
a single Gaussian in \mES and a Crystal Ball function~\cite{ref:CB} in
\mDz; (2) combinatorial 
background with a real \Dz: an Argus~\cite{ref:ARGUS} function in \mES and a
 Crystal Ball function in \mDz; (3) combinatorial background with
 a combinatorial \Dz: an Argus 
function in \mES and a first order polynomial in \mDz.
 The Crystal Ball parameters for \mDz in
components (1) and (2) share the same values, as does the Argus
parameter in (2) and (3). 

\begin{figure}[htb]
\begin{center}
\includegraphics[width=0.9\textwidth]{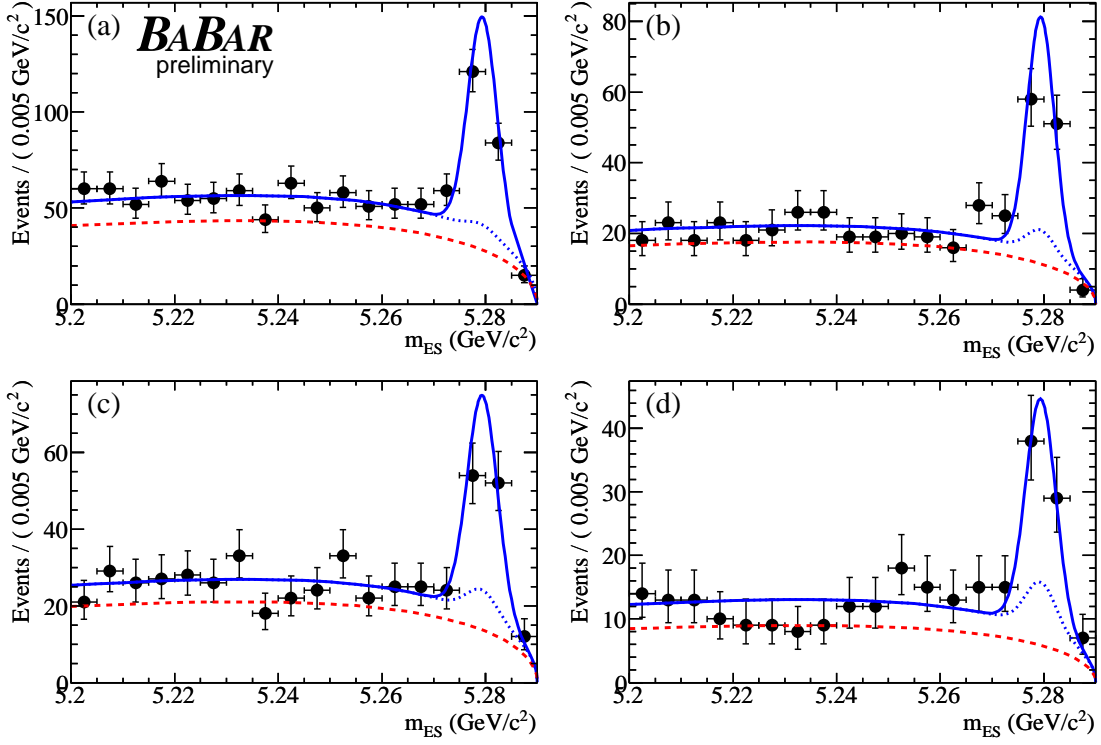}
\end{center}
\caption{Distributions of \mES for (a) $\Dz\piz$, (b)
  $\Dz\eta^{(\prime)}$, (c) $\Dz\omega$ and (d) $\Dstarz\piz,\eta$ for
  events with 
  \mDz within 14~\mevcc of the nominal value. Curves are: (solid)
  overall PDF projection; (dotted) background (including peaking) PDF;
  (dashed) 
  contribution from background with fake \Dz. }
\label{fig:yield}
\end{figure}

\begin{figure}[htb]
\begin{center}
\includegraphics[width=0.9\textwidth]{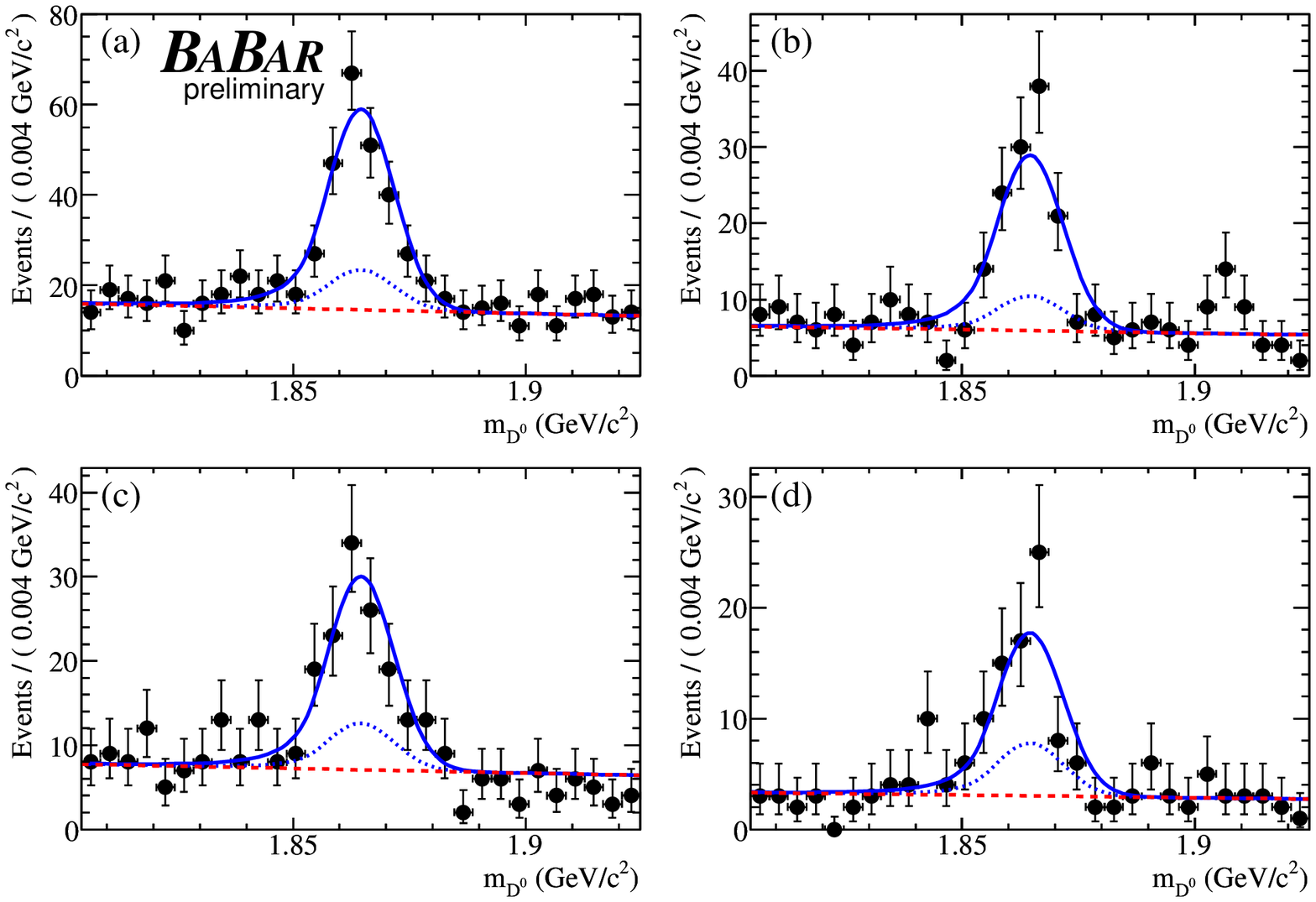}
\end{center}
\caption{Distributions of \mDz  for (a) $\Dz\piz$, (b)
  $\Dz\eta^{(\prime)}$, (c) $\Dz\omega$ and (d) $\Dstarz\piz,\eta$
  for events with
  $\mES>5.27~\gevcc$. Curves are: (solid)
  overall PDF projection; (dotted) background (including peaking) PDF;
  (dashed) 
  contribution from background with fake \Dz. }
\label{fig:yieldD}
\end{figure}

\begin{figure}[htb]
\begin{center}
\includegraphics[width=0.98\textwidth]{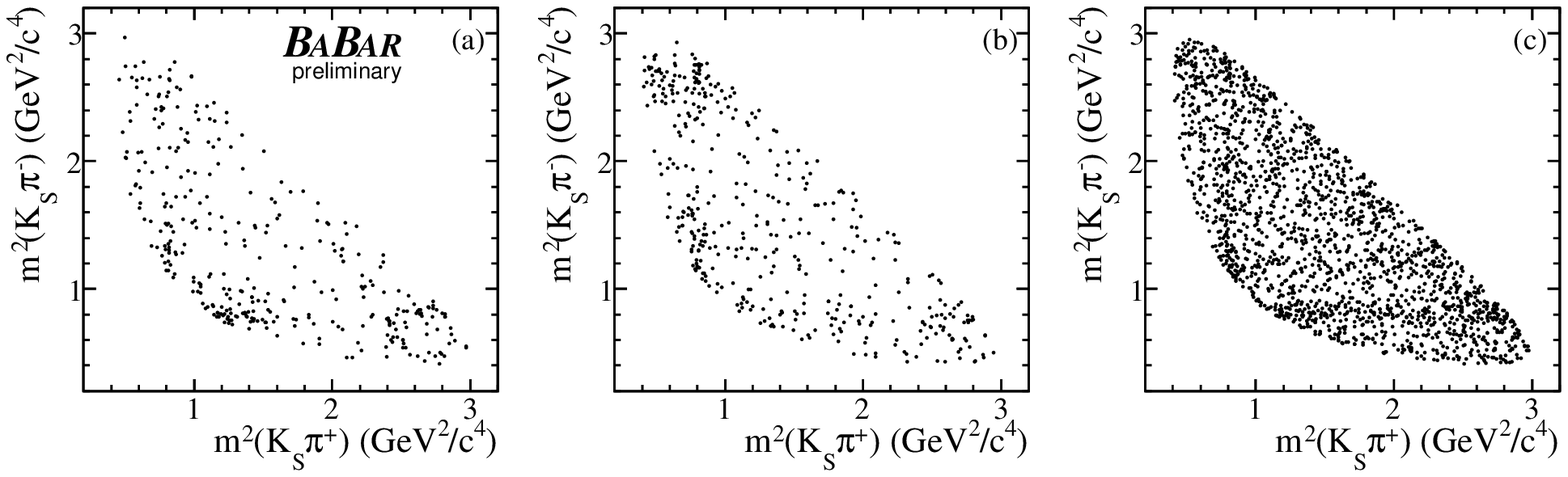}
\end{center}
\caption{Dalitz distributions for (a) \Bz-tagged and (b)
  \Bzb-tagged events in the signal 
  region, $\mES>5.27~\gevcc$,
  and (c) events in the \mES sideband $\mES<5.26~\gevcc$. In all
  cases, the \Dz mass is required to be within 20~\mevcc of the nominal value.}
\label{fig:datadalitz}
\end{figure}

The results of the fit are shown in Figures~\ref{fig:yield}
and~\ref{fig:yieldD}; the 
yields are shown in Table~\ref{tab:yield}. We merge the $\Dz\eta$
and $\Dz\etapr$ samples, as well as the $\Dstarz\piz$ and $\Dstarz\eta$
samples.
The Dalitz distributions in the signal region and \mES sideband are shown
in Figure~\ref{fig:datadalitz}.
There are irreducible
backgrounds that peak in both \mES and \mDz, which cannot be
discriminated against with our PDF. The majority of this peaking
background is from $\Bp\ra D^{(*)0}\rho^+$. We estimate the amount of
the peaking background using the simulated generic Monte Carlo
samples. The number of signal events after subtracting the peaking
background is shown in the last row in Table~\ref{tab:yield}.

\begin{table}[htb]
\caption{Signal event yields.}
\begin{center}
\begin{tabular}{lccccc}
\hline\hline
Decay mode & $\Dz\piz$ & $\Dz\eta,\eta'$ & $\Dz\omega$ &
$\Dstarz\piz,\eta$ & Total \\
\hline
Raw peak yield & $175 \pm 17$ & $97 \pm 11$ & $93 \pm 12$ &
$59 \pm 9$ & $424 \pm 25$ \\
\hline
Peaking background subtracted yield & $168 \pm 19$ & $87 \pm 12$ & $82 \pm 13$ & $47 \pm 9$ & $384 \pm 28$ \\
\hline\hline
\end{tabular}
\end{center}
\label{tab:yield}
\end{table}

%% file: dalitzmodel.tex
\section{DALITZ PLOT MODEL}
\label{sec:dalitz}

The \Dz decay amplitude is determined from an unbinned
maximum-likelihood fit to a high-purity sample of
$\Dstarp\ra\Dz\pip$ decays. We use the isobar formalism described
in~\cite{isobarCLEO} to express $A_D$ as a sum of two-body decay 
matrix element (subscript $r$) and a non-resonant (subscript NR)
contribution,
\begin{equation}
A_D(m_+^2,m_-^2) = \sum_r a_r e^{i\phi_r} A_r(m_+^2,m_-^2) +
a_\mathrm{NR}e^{i\phi_\mathrm{NR}}\,,
\end{equation}
where each term is parameterized with an amplitude $a_r$ and a phase
$\phi_r$. The function $ A_r(m_+^2,m_-^2)$ is the Lorentz-invariant
expression for the matrix element of a \Dz meson decaying into
$\KS\pip\pim$ through an intermediate resonance $r$, parametrized as a
function of the position in the Dalitz plane.

The resonances in the model for \Dz are: (a) $[\KS\pim]$ resonances:
$\Kstar(892)^-$, $\Kstar_0(1430)^-$,
 $\Kstar_2(1430)^-$, $\Kstar(1410)^-$,
and $\Kstar(1680)^-$; (b) $[\KS\pip]$ resonances (doubly-Cabibbo
suppressed): $\Kstar(892)^+$, 
$\Kstar_0(1430)^+$, and $\Kstar_2(1430)^+$;
(c) $[\pip\pim]$ resonances: $\rho(770)$, $\omega(782)$, $f_0(980)$,
$f_0(1370)$, $f_2(1270)$, $\rho(1450)$ and two scalar resonances
$\sigma$ and $\sigma^\prime$.  For $\rho(770)$ and $\rho(1450)$ we use
the functional form suggested in~\cite{GJGJJS}, while the remaining
resonances are parametrized by a spin-dependent relativistic
Breit-Wigner distribution. The means and widths of the resonances are
taken from the PDG~\cite{ref:PDG}, except for $\sigma$ and
$\sigma^\prime$, which are obtained from the Dalitz plot fit. The inclusion
of $\sigma$ and $\sigma^\prime$ significantly improves the Dalitz plot
fit quality.
However, since these two scalars are not well established, we
consider the systematic effect of using a model without them. 
More details about the Dalitz plot model and parameters
can be found in~\cite{babar_gamma}. 
We neglect the detector resolution in the Dalitz plot model because the
resolution ($\simeq 4~(\mevcc)^2$) is much smaller than the resonance
widths. Only $\omega(782)$ has an intrinsic width comparable to the
mass resolution. We increase its width artificially in the systematic
study.

%% file: analysis.tex
\section{TIME-DEPENDENT ANALYSIS}
\label{sec:analysis}

We model the time-dependent Dalitz plot distribution in a
PDF that consists of four components: signal, background with a real
\Dz, background with a fake \Dz and background that peaks in both \mES
and \mDz. The $(\mES,\mDz)$ model for the first three components has
been described in Section~\ref{sec:selection}. The peaking background
component shares the same $(\mES,\mDz)$ shape with the signal component.
The background fractions are determined from
a fit to $(\mES,\mDz)$ distributions and from generic Monte Carlo
samples (for peaking background) for each \Bz mode group and each
tagging category. Each event is assigned signal and background
probabilities based on the two-dimensional PDF.

The time-dependent Dalitz model for signal is based on
Equation(\ref{eq:Probpm}). We modify the equation to take into account
mistagging and imperfect \dt reconstruction, following the methods
used in our other time-dependent analyses~\cite{tagging}, i.e., 
an additional factor $(1-2 w)$ is added to the
$\cos(\dm\dt)$ and $\sin(\dm\dt)$ terms, and the equation is convolved
with a  three-Gaussian \dt resolution function. There are six tagging
categories with different mistag fractions $w$. We also allow
the $w$ of each category to be different for \Bz and \Bzb tags.
The means and widths of the core and the
second Gaussian are parameterized with scale factors multiplied by
\sigmadt. The mean and width of the third (outlier) Gaussian are fixed
at 0~ps and 8~ps, respectively. The mistag rates and the resolution
function are determined from control samples of $\Bz\ra
D^{(*)}\pi,\rho, a_1 $ decays. Most of the resolution function
parameters are consistent among the six tagging
categories, except for the core Gaussian mean and scale
factor, where the Lepton tagged sample is significantly different from
others. We allow these two parameters to be different for Lepton tag.

The model for the background with a fake \Dz is determined from the
\Dz sideband data. The \dt model consists of a prompt component and a
exponential decay component with an effective lifetime. The resolution
function is a Gaussian whose mean and width are scaled by \sigmadt,
plus an outlier Gaussian. The mean of the core Gaussian and
the fraction of the prompt component are allowed to be different
between the Lepton tag and the other tags. 

The Dalitz distribution for background is modeled by an incoherent
mixture of several 
resonances and a phase-space distribution,
\begin{equation}
\label{eq:dalitzbkg}
{\cal P}(m_+^2,m_-^2) = |a_\mathrm{NR}|^2 + \sum_r |a_r|^2 |A_r(m_+^2,m_-^2)|^2
\;.
\end{equation}
We find that the model describes the \Dz sideband data well if we
include $\Kstar(892)^-$, $\Kstar(892)^+$, 
$\Kstar_0(1430)^-$, $\rho(770)$, $\rho(1450)$ and $\sigma$ resonances
in the model. We also check that the Dalitz distribution is
independent of the tagging category, the flavor tag, and \dt.

Based on a study using the generic Monte Carlo samples, the background
with a real 
\Dz comes mostly from \ccbar continuum events. We therefore model the
\dt distribution with a prompt component convolved with a core Gaussian plus
outlier resolution function. The parameters are determined from the
generic Monte Carlo sample. 

The Dalitz model for this background is either $A_D(m_+^2,m_-^2)$ or
$A_{\Dbar}(m_+^2,m_-^2)$ based on the flavor of the tagging side \Btag. 
If \Btag is tagged as \Bz (\Bzb), the $D$ in the reconstructed
candidate is more likely to be a \Dz (\Dzb).
Since they are not \BB events, the mistag rates are not the same as
those for the signal. However, we don't have reliable mistag values
for continuum events. We therefore use the mistag rates for signal in the
nominal fit and vary them to estimate the systematic uncertainty.

The peaking background being mostly from charged \B decays, the \dt model is an
exponential decay with the lifetime fixed at the
\Bp lifetime. 
The Dalitz model is identical to that for
combinatorial background with real \Dz, except that the mistag rates
can be different. Again we fix the mistag rates to those for the signal, and
vary them for systematic uncertainty.

\begin{figure}[htb]
\begin{center}
\includegraphics[width=0.98\textwidth]{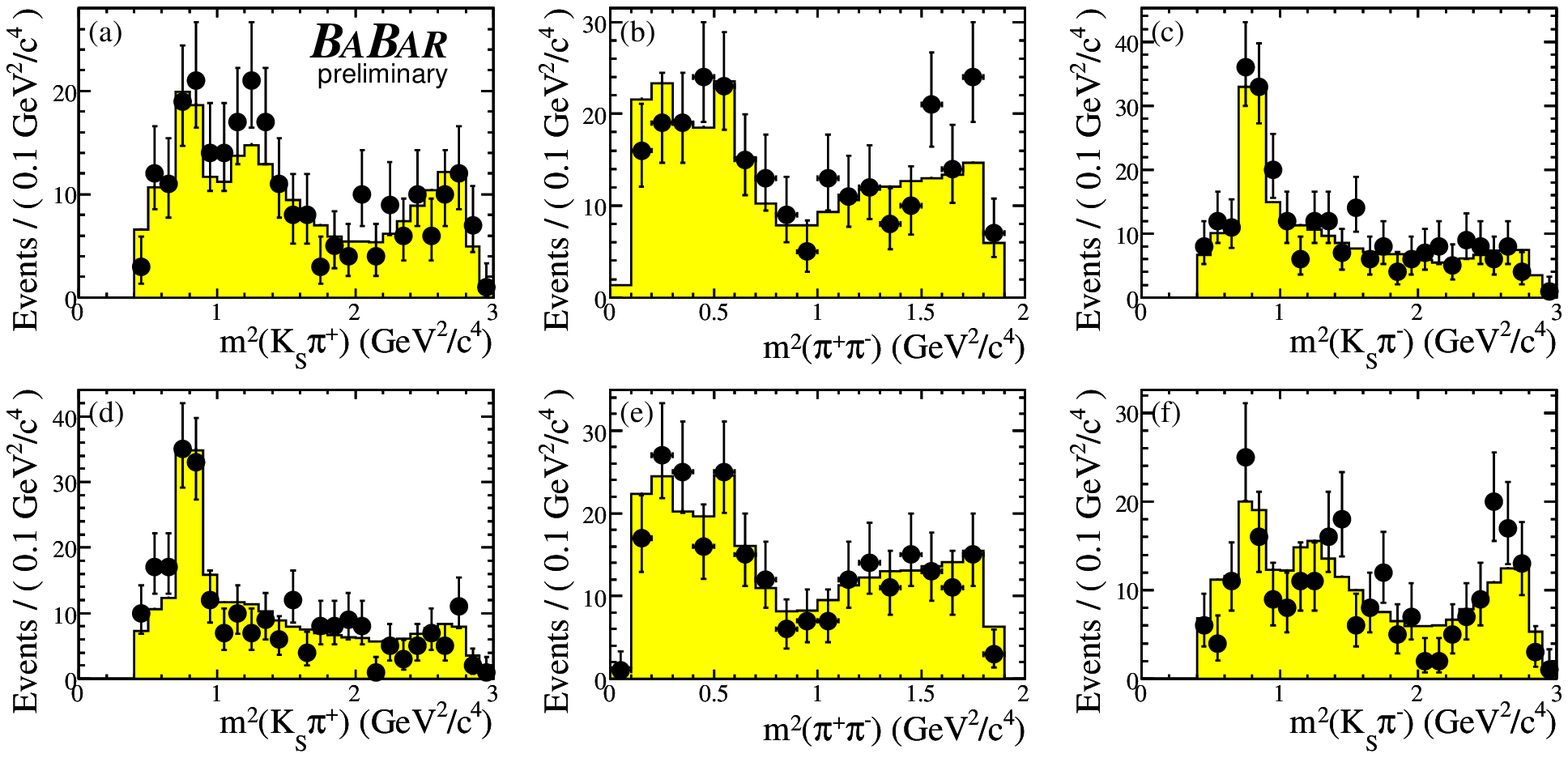}
\end{center}
\caption{Projection on (a,d) \msqKsp, (b,e) \msqpm, and (c,f) \msqKsm
  for (a,b,c) 
  \Bz-tagged and (d,e,f) \Bzb-tagged events separately, in the signal region
  ($\mES>5.27~\gevcc$, $|\mDz-\mDz^\mathrm{PDG}|<20~\mevcc$).
  Points with error bars are data; histograms are from the PDF.}
\label{fig:datatoyDproj1}
\end{figure}

\begin{figure}[htb]
\begin{center}
\includegraphics[width=0.5\textwidth]{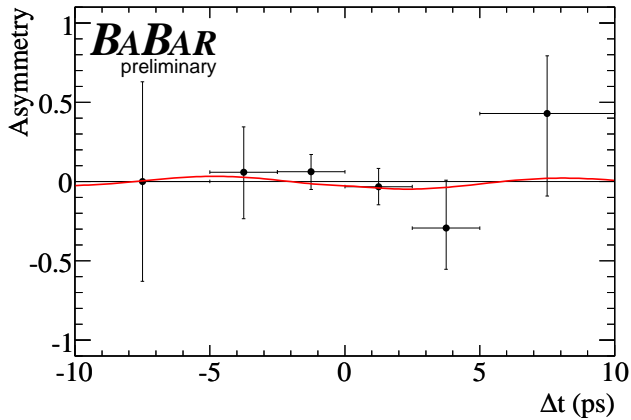}
\end{center}
\caption{Asymmetry distribution for the events in
  $\Dz\ra\KS\rho(770)$ region. The curve is the result of the
  PDF.} 
\label{fig:d0rhoasym}
\end{figure}

In the nominal fit, we allow \cosbb, \sinbb and \abslambda to float
and fit to all data samples and tagging categories simultaneously. 
The \Bz lifetime and mixing frequency are fixed at the PDG values.
The fit results are shown in Table~\ref{tab:result}, where the result
for which the \sinbb is fixed at the world average and \abslambda at
one is also included. We also allow the 
\mES shape and 
background fractions to float in the fit. We find no significant
difference in either the central values or the statistical
uncertainties. The projections on the Dalitz plot variables are shown in
Figure~\ref{fig:datatoyDproj1} and are compared with the distributions
described by the model. Figure~\ref{fig:d0rhoasym} shows the
time-dependent \CP asymmetry for events in $\Dz\ra\KS\rho(770)$ region
($|m(\pip\pim)-0.77|<0.25~\gevcc$), where the \CP asymmetry is
expected to be enhanced.
The apparent asymmetry in Figure~\ref{fig:d0rhoasym} is small compared
to \sinbb due to 
the dilution factors from mistagging, background and contributions
from non-resonance and resonances other than $\rho$.

\begin{table}[htb]
\caption{Results of the fits to data. Errors are statistical only.}
\begin{center}
\begin{tabular}{lccc}
\hline\hline
Final state & \cosbb & \sinbb & \abslambda \\
\hline
$\Dz\piz$ & $1.1^{+0.8}_{-0.9}$ & $1.0\pm0.5$ &
$1.13^{+0.17}_{-0.14}$ \\
$\Dz\eta^{(\prime)}$ & $0.4\pm1.1$ & $-0.1^{+0.9}_{-1.0}$ &
$0.96^{+0.19}_{-0.16}$ \\ 
$\Dz\omega$ & $-0.4^{+1.3}_{-1.4}$ & $0.7\pm 1.0$ &
$0.61^{+0.17}_{-0.15}$ \\ 
$\Dstarz\piz/\eta$ & $0.3\pm1.4$ & $-0.8^{+1.0}_{-0.9}$ &
$1.05^{+0.35}_{-0.25}$ \\ 
\hline
All & $0.54\pm0.54$ & $0.45\pm0.35$ & $0.98\pm 0.09$ \\
All & $0.55 \pm 0.52$ & $0.685$ (fixed) & 1 (fixed) \\
\hline\hline
\end{tabular}
\end{center}
\label{tab:result}
\end{table}

%% file: systematic.tex
\section{SYSTEMATIC STUDIES}
\label{sec:systematics}

Dependence on the choice of Dalitz plot model in signal
is expected to be one of the largest
systematic uncertainties. We estimate the systematic uncertainty by
comparing the nominal Dalitz model and an alternative model where the
two scalar resonances $\sigma$ and $\sigma^\prime$ are removed. 
 We generate 300 toy datasets using the
parameters from the nominal fit; each
toy dataset has 50 times the data size. We fit to each toy dataset
using both nominal Dalitz model and the alternative Dalitz model. 
 We find that both \sinbb and \cosbb shift
significantly. We take the quadratic sum of the mean and RMS of
the distribution of the difference between the two models as the
systematic uncertainty due to Dalitz model uncertainty.
The detector resolution on the Dalitz plot is neglected. Only
$\omega(782)$ has an intrinsic width comparable to the 
mass resolution. We increase its width from 8.5~\mev to 10~\mev and
find no significant change in the results.

As described in Equation(\ref{eq:dalitzbkg}), the
background Dalitz model is described by an ad hoc incoherent mixture of
several resonances and a phase-space distribution.
Alternatively, we use a background model containing only \Kstar and
a phase-space distribution to describe the \Dz sideband Dalitz
distribution. This model describes the data rather poorly. However, the
changes to the final results are quite small.

We vary the \Bz lifetime and mixing by their uncertainty quoted in the
PDG~\cite{ref:PDG2006},
and other fixed parameters by their statistical uncertainty in the
fits to control samples, to estimate the systematic uncertainties. The
parameters include the amplitudes and phases of the Dalitz model,
\mES, \mDz and background \dt distributions, background
fractions, mistag rates and resolution functions. In addition, we vary
several resolution function parameters that were fixed in the fit to
the control sample: outlier bias from $-2$ to $+2$~ps, outlier width
from 4 to 12~ps, and the second Gaussian scale factor from 2 to 5. 

The mistag rates in both peaking and combinatorial background with a
real \Dz are the same as the signal mistag rates in the nominal
fit. We vary the mistag rate of each tagging category by $\pm30\%$ for
both backgrounds to estimate the systematic uncertainties.

We fit a two-dimensional third-order polynomial to signal Monte Carlo
samples 
to parametrize the reconstruction efficiency variation over the Dalitz
plot.  We vary
the parameters by their one-sigma statistical uncertainty and the
variation in the final answer is again negligible.
If we simply assume the efficiency is a constant across the Dalitz
plot, the changes to the results are still quite small and we treat these
differences as systematic uncertainties.
The systematic uncertainties are summarized in Table~\ref{tab:syst-sum}.

\begin{table}[htb]
\caption{Summary of systematic uncertainties.}
\begin{center}
\begin{tabular}{lccc}
\hline\hline
Item & \cosbb & \sinbb & \abslambda\\
\hline
Signal Dalitz model      & 0.184 & 0.073 & 0.002 \\
\hline
Signal Dalitz parameters       & 0.068 & 0.026 & 0.006 \\
Background Dalitz parameters         & 0.002 & 0.001 & 0.002 \\
Background Dalitz model        & 0.004 & 0.006 & 0.008 \\
$\dm$                    & 0.003 & 0.002 & 0.000 \\
$\tau_{\Bz}$             & 0.003 & 0.001 & 0.000 \\
$\tau_{\Bp}$             & 0.003 & 0.000 & 0.000 \\
Mistag, resolution, etc. & 0.043 & 0.043 & 0.002 \\
Peaking background fraction    & 0.020 & 0.018 & 0.005 \\
Mistag in combinatorial background with \Dz& 0.002 & 0.001 & 0.001 \\
Mistag in peaking background   & 0.002 & 0.001 & 0.001 \\
Dalitz plot efficiency   & 0.002 & 0.001 & 0.000 \\
\hline
Total (non-Dalitz-model) & 0.083 & 0.054 & 0.012 \\
\hline
Total                    & 0.202 & 0.091 & 0.012 \\
\hline\hline
\end{tabular}
\end{center}
\label{tab:syst-sum}
\end{table}

%% file: result.tex
\section{RESULTS}
\label{sec:physics}

We have measured the CKM phase \cosbb and \sinbb
using a time-dependent Dalitz plot analysis of
$\Dz\ra\KS\pip\pim$ decays in $\Bz\ra\Dz\hz$ decays. We obtain
\begin{eqnarray}
\cosbb &= & 0.54 \pm 0.54 \pm 0.08 \pm 0.18 \\
\sinbb &= & 0.45 \pm 0.35 \pm 0.05 \pm 0.07 \\
\abslambda &=& 0.975^{+0.093}_{-0.085} \pm 0.012 \pm 0.002 \;,
\end{eqnarray}
where the first error is
statistical, the second is experimental systematic uncertainty, and
the third is the signal Dalitz plot model uncertainty.
The statistical correlation between \cosbb and \sinbb is 7\%, and less
than 5\% between \abslambda and \cosbb or \sinbb.
The result is consistent with $\sinbb_0= 0.685\pm 0.032$ and $\cosbb_0=
\pm\sqrt{1-\sin^{2}2\beta_0}= \pm 0.729$ obtained from high precision
measurement using \Bz to charmonium \Kz decays, and consistent with no \CP
violation in \B decay ($\abslambda =1$). If \sinbb is fixed at 0.685
and \abslambda at one in our analysis, we obtain
\begin{equation}
\cosbb = 0.55 \pm 0.52\pm 0.08 \pm 0.18 \;.
\end{equation}

\begin{figure}[htb]
\begin{center}
\includegraphics[width=0.9\textwidth]{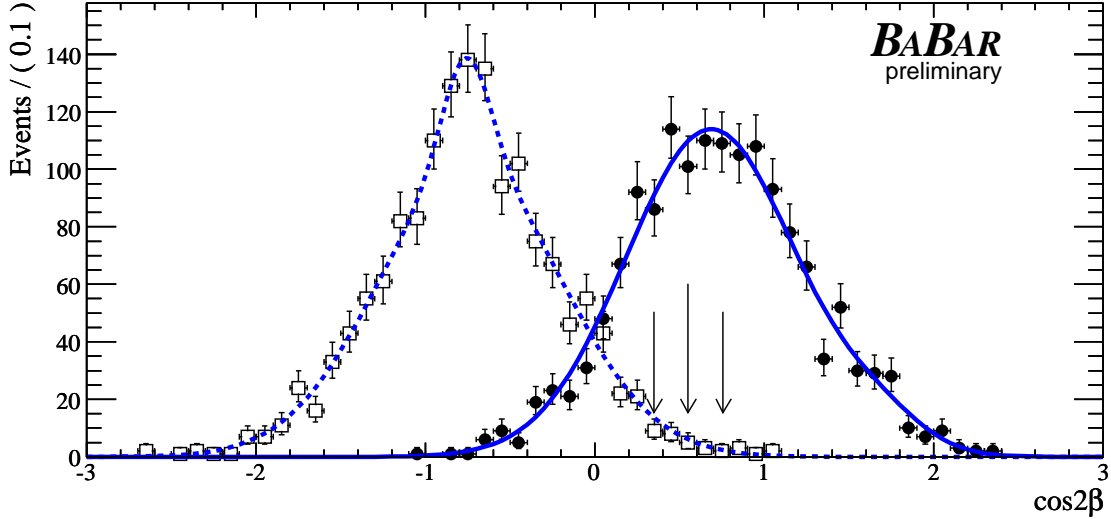}
\end{center}
\caption{Distribution of \cosbb obtained from two sets of 1500 simulated
  experiments of the same size as the data sample, as described in the
  text. Distribution in solid dots (open squares) is for samples
  generated with $\cosbb= |\cosbb_0|$ ($-|\cosbb_0|$). Solid (dashed)
  curve is the corresponding two-Gaussian function $h_+(x)$ ($h_-(x)$).
  Three vertical arrows indicate the
  central value of \cosbb and plus/minus systematic uncertainty. } 
\label{fig:confid}
\end{figure}

This result allows one to distinguish the two possible solutions of
angle $2\beta_0$. We generate 1500 parametrized simulation samples of
the same size as the data sample, where we use $\sinbb=\sinbb_0$, $\cosbb=
|\cosbb_0|$ and $\abslambda=1$. We then fit to each sample
with \cosbb as the only free parameter, and use a two-Gaussian
function $h_+(x)$ to fit to the distribution of the 1500
results. We repeat the same exercise using $\cosbb= -|\cosbb_0|$ to
generate simulation samples 
and obtain another two-Gaussian
distribution $h_-(x)$. The distributions are shown in
Figure~\ref{fig:confid}. We define the confidence level (CL) at which the 
$\cosbb= -|\cosbb_0|$ solution is excluded when we
observe $\cosbb = x$ as $h_+(x)/[h_+(x)+h_-(x)]$. 
We calculate the CL for $\cosbb= 0.55$, $0.35$ and $0.75$ to account
for the systematic uncertainty and use the smallest value, $87\%$, as
the final result.

\section{CONCLUSIONS}
\label{sec:summary}

We have studied the time-dependent Dalitz distribution in $\Bz\ra
D^{(*)0}\hz$ decays and determined the \CP asymmetry parameters
\sinbb, \cosbb and \abslambda. The results are consistent with the
Standard Model expectations.
Assuming \sinbb is equal to $\sinbb_0$ found 
in \Bz to charmonium \Kz analyses and no \CP violation in \B decays,
we determined that the solution $\cosbb=
-\sqrt{1-\sin^{2}2\beta_0}$ is excluded at an $87\%$ confidence level.

%% file: pubboard/acknowledgements.tex
We are grateful for the 
extraordinary contributions of our \pep2\ colleagues in
achieving the excellent luminosity and machine conditions
that have made this work possible.
The success of this project also relies critically on the 
expertise and dedication of the computing organizations that 
support \babar.
The collaborating institutions wish to thank 
SLAC for its support and the kind hospitality extended to them. 
This work is supported by the
US Department of Energy
and National Science Foundation, the
Natural Sciences and Engineering Research Council (Canada),
Institute of High Energy Physics (China), the
Commissariat \`a l'Energie Atomique and
Institut National de Physique Nucl\'eaire et de Physique des Particules
(France), the
Bundesministerium f\"ur Bildung und Forschung and
Deutsche Forschungsgemeinschaft
(Germany), the
Istituto Nazionale di Fisica Nucleare (Italy),
the Foundation for Fundamental Research on Matter (The Netherlands),
the Research Council of Norway, the
Ministry of Science and Technology of the Russian Federation, and the
Particle Physics and Astronomy Research Council (United Kingdom). 
Individuals have received support from 
the Marie-Curie IEF program (European Union) and
the A. P. Sloan Foundation.